\documentclass[sigconf, nonacm]{acmart}
\usepackage{enumitem}
\usepackage{subcaption}
\usepackage{cleveref}
\definecolor{firebrick}{HTML}{B22222} 
\definecolor{clightgray}{rgb}{0.83, 0.83, 0.83}
\newcommand{\styledtextcircled}[3]{
  \tikz[baseline=(char.base)]{
    \node[shape=circle, draw=#1, 
    line width=0.2mm, 
    inner sep=0.6pt] (char) {#3};
  }
}
\newcommand{\cir}[1]{
  \styledtextcircled{firebrick}{clightgray}{\textbf{#1}}
}

\newcounter{mycodeboxcounter}

\usepackage[most]{tcolorbox}    	
\newtcolorbox{mycodebox}[1][]{
    standard jigsaw,
    opacityback=0,  
    boxrule = 0.pt,
    left=0pt,
    right=0pt,
    top=0pt,
    bottom=0pt,
    sharp corners, 
    width=\linewidth,      
    halign=flush left,     
    before skip=5pt,       
    after skip=5pt,        
    grow to right by=0mm,   
}
\usepackage{fancyvrb}
\newcommand{\cteal}[1]{\textcolor{teal}{#1}}
\newcommand{\cpurple}[1]{\textcolor{purple}{#1}}
\newcommand{\vteal}[1]{\textcolor{teal}{\texttt{#1}}}
\newcommand{\vpurple}[1]{\textcolor{purple}{\texttt{#1}}}
\newcommand{\vtextbf}[1]{\textbf{{\texttt{#1}}}}

\usepackage{xcolor}
\usepackage{listings}
\definecolor{codegray}{rgb}{0.5,0.5,0.5}
\definecolor{codeblue}{rgb}{0,0,1}
\definecolor{codegreen}{rgb}{0,0.5,0}
\definecolor{codepurple}{rgb}{0.58,0,0.82}
\definecolor{codered}{rgb}{0.7, 0.1, 0.2}
\definecolor{lightblue}{RGB}{230,240,255}
\lstdefinestyle{mystyle}{
    backgroundcolor=\color{gray!3},
    commentstyle=\color{codegray},
    keywordstyle=\bfseries\color{black},
    numberstyle=\footnotesize\color{gray},
    stringstyle=\color{codepurple},
    basicstyle=\ttfamily\footnotesize,
    breakatwhitespace=false,         
    breaklines=true,                 
    captionpos=b,                    
    keepspaces=true,                 
    numbers=left,                    
    numbersep=5pt,                  
    showspaces=false,                
    showstringspaces=false,
    showtabs=true,                  
    tabsize=2,
    frame=single,
    framerule=0.1pt,
    rulecolor=\color{gray},
    xleftmargin=5pt,
    xrightmargin=5pt
  }
\newcommand\vldbdoi{XX.XX/XXX.XX}
\newcommand\vldbpages{XXX-XXX}
\newcommand\vldbvolume{14}
\newcommand\vldbissue{1}
\newcommand\vldbyear{2020}
\newcommand\vldbauthors{\authors}
\newcommand\vldbtitle{\shorttitle} 
\newcommand\vldbavailabilityurl{URL_TO_YOUR_ARTIFACTS}
\newcommand\vldbpagestyle{plain}

\begin{document}
\title{
Revisiting Page Migration In Main-Memory Databases For Modern Hardware
}

\author{Yeasir Rayhan and Walid G. Aref}
\affiliation{
  \institution{Purdue University, West Lafayette, IN, USA}
  \city{}
  \country{}
}
\email{{yrayhan, aref}@purdue.edu}

\begin{abstract}
Modern hardware architectures, e.g., NUMA servers, chiplet processors, tiered and disaggregated memory systems have significantly improved the  performance of Main-Memory Databases, and are poised to deliver further improvements in the future. However, realizing this potential depends on the database system's ability to efficiently migrate pages among different NUMA nodes, and/or memory chips as the workload evolves. Modern Main-Memory Databases
offload the migration procedure to the operating system without accounting for the workload and its migration characteristics. In this paper, we propose a custom system call \vpurple{move\_pages2} as an alternate to Linux's own \vpurple{move\_pages} system call. 
{In contrast to the original move\_pages, move\_pages2 allows partial migration and exposes two configuration knobs, enabling a Main-Memory Database tailor the migration process to its specific requirements. Experiments on a main-memory B$^+$-Tree for a YCSB-like workload show that the proposed \texttt{move\_pages2} custom system call 
improves the B$^+$-Tree query throughput by up to 2.3$\times$ 
and migrates 
up to 2.6$\times$ more memory pages, outperforming the native Linux system call on modern hardware architectures.}
\end{abstract}

\maketitle

\pagestyle{\vldbpagestyle}
\begingroup\small\noindent\raggedright\textbf{PVLDB Reference Format:}\\
\vldbauthors. \vldbtitle. PVLDB, \vldbvolume(\vldbissue): \vldbpages, \vldbyear.\\
\href{https://doi.org/\vldbdoi}{doi:\vldbdoi}
\endgroup
\begingroup
\renewcommand\thefootnote{}\footnote{\noindent
This work is licensed under the Creative Commons BY-NC-ND 4.0 International License. Visit \url{https://creativecommons.org/licenses/by-nc-nd/4.0/} to view a copy of this license. For any use beyond those covered by this license, obtain permission by emailing \href{mailto:info@vldb.org}{info@vldb.org}. Copyright is held by the owner/author(s). Publication rights licensed to the VLDB Endowment. \\
\raggedright Proceedings of the VLDB Endowment, Vol. \vldbvolume, No. \vldbissue\ %
ISSN 2150-8097. \\
\href{https://doi.org/\vldbdoi}{doi:\vldbdoi} \\
}\addtocounter{footnote}{-1}\endgroup


\section{Introduction}
Main-Memory Databases (MMDB, for short)~\cite{FaerberKLLNP17} have emerged as a promising alternative to conventional disk resident databases in order to mitigate the performance bottlenecks associated with disk I/O. By storing the entire database in main memory, an MMDB eliminates the need to access disks altogether. Despite early efforts, e.g.,~\cite{GawlickK85,LehmanC86vldb,LehmanC86sigmod,LehmanC87}, MMDBs have gained widespread traction only in recent years, mainly due to the decreasing cost and increasing capacity of main memory. This has led to full-fledged systems in both academia (e.g., H-Store~\cite{StonebrakerMAHHH07,KallmanKNPRZJMSZHA08}, HyPeR~\cite{0001MK15}, Silo~\cite{TuZKLM13}, Hyrise~\cite{GrundKPZCM10,DreselerK0KUP19}
), and industry (e.g., Hekaton~\cite{DiaconuFILMSVZ13}, VoltDB~\cite{StonebrakerW13}, Oracle TimesTen~\cite{LahiriNF13}, SAP HANA~\cite{SikkaFGL13}). 
These MMDBs are built upon legacy hardware, based on the primary assumptions that the memory access latency is uniform and 
that
the memory is tightly coupled to the compute cores. Page migration holds little to no value in these legacy hardware architectures, and as a result has largely been neglected in MMDBs until now.

{
However, in the past decade, the hardware landscape has evolved rapidly with the introduction of NUMA architectures, chiplet-based processor designs~\cite{amd_zen2,amd_zen3,amd_zen4,intel_sierra_forest}, and emerging interconnect technologies, e.g., CXL~\cite{CXL}, OpenCAPI~\cite{ocapi_cxl}, and CCIX~\cite{ccix}. These hardware developments invalidate the earlier assumptions of uniform memory access latency and tight coupling of memory and compute chips. In NUMA- and chiplet-based architectures,  memory access latency is non-uniform. On top of that, the introduction of CXL-like interconnect technology has introduced memory configurations where memory is loosely coupled, i.e., a tiered-memory architecture, and is even decoupled, i.e., a disaggregated memory architecture, from the compute chip. The dominant theme underlying these new hardware advancements is the introduction of heterogeneity in hardware, and thus leading to variable access latencies across different hardware chips. For example, access to remote-socket memory in a NUMA server can be $4\times$ more expensive than accessing local-socket memory~\cite{BangMPB22,BangMPB20}. In a CXL-enabled tiered memory architecture, accessing CXL memory can be $5\times$ more expensive than accessing local DRAM memory~\cite{LiuHWBNJNL25}. 
}

{
For optimal performance on a NUMA server, data must be strategically placed close to the compute core that initiates the data request. This minimizes cross-socket data movement. Additionally, as the workloads evolve and the working sets shift, data must be continuously moved around among the NUMA nodes in an online manner to sustain optimal DBMS performance. For example, SAP-HANA~\cite{sap_hana}, a commercial main-memory column-store, demonstrates that an adaptive data-placement strategy~\cite{PsaroudakisSMSA16} can yield up to $4\times$ performance improvement in terms of query throughput. The same principle also applies to tiered memory architectures. Data needs to be continuously moved around so that hot data resides in the closest memory, i.e., local DRAM, and cold data resides in the farthest memory, i.e., CXL memory~\cite{HaoZYS24}. }

{To support this dynamic relocation of data, modern MMDBs~\cite{HaoZYS24,RiekenbrauckWLR24,FogliPG24} often rely on the support of the operating system in the form of {on-demand} page migration that is the de facto standard for moving memory pages across NUMA nodes and volatile memory tiers. Thus, the efficiency of page migration is critical to achieving optimal performance in modern hardware architectures. For instance, the three-tier buffer pool manager~\cite{RiekenbrauckWLR24} comprising DRAM, CXL Memory, and Disk in Hyrise~\cite{DreselerK0KUP19}, identifies page migration as a key bottleneck in a tiered buffer pool, that can cause up to 2$\times$ performance degradation. This further validates 
the
importance of page migration in 
a Main-Memory Database in the context of modern hardware, and highlights the need for greater attention from the research community as we enter the next phase of main-memory database design for modern hardware. 
}

Linux~\cite{linux_git} handles page migration 
through the system call \vpurple{move\_pages}~\cite{linux_move_pages}. \texttt{move\_pages}
accepts as input a list of page addresses along with 
their
corresponding destination NUMA nodes (memory tiers). It alters the physical address of a page by moving it to its destination NUMA node (memory tier) without changing the page's virtual address. A typical page migration involves several key steps. \cir{1}~The user space application invokes a
\verb|move_pages| system call, and traps to the kernel to handle the 
migration; \cir{2}~The page is locked, and the page table entry (PTE, for short, is an entry that maps a virtual address to a physical address in a process's page table) of the migrating page is marked as {\em a migration entry} to ensure exclusive access during migration; \cir{3}~A TLB (short for Translation Lookaside Buffer that caches recent virtual-to-physical address translations) shootdown is issued to invalidate all 
PTEs of the migrating page across the TLBs of all the computing cores; \cir{4}~A new page is created, and is locked; \cir{5}~The contents of the old page are copied to the new page; \cir{6}~The page locks are dropped from both the old and new pages, and \cir{7}~The 
corresponding PTE is remapped to point to the new page. Section~\ref{sec:sys_call} details these steps.

The efficiency of page migration is critical to an MMDB's performance especially in a dynamic multi-threaded environment, where the workload is continuously changing, and the memory pages need to be moved around accordingly, while keeping the database online. An efficient implementation of the \verb|move_pages| system call ensures that the migration process is fast, and the interference among the kernel threads and the MMDB worker threads is kept to a minimum during migration, thus reducing hardware resource contention. However, 
page migration in today’s systems has high overhead and is inefficient, e.g., see~\cite{YanLNB19, XiangLD0RY024}. Recent efforts to improve the performance of the \verb|move_pages| system call have  focused mainly on reducing the overheads from both the hardware~\cite{AwadBBSL17,Amit17,GandhiKACHMNSU16,CoxB17} and the software~\cite{YanLNB19, XiangLD0RY024} sides.

In this paper, we follow a different approach complementary to those in~\cite{AwadBBSL17,Amit17,GandhiKACHMNSU16,CoxB17,YanLNB19, XiangLD0RY024}. We implement a custom system call termed \vpurple{move\_pages2} in Linux 6.8.0. {The design of this new system call is based on two key principles: 1)~Delegating control over the migration process to the MMDB as much as possible, and 2)~Enabling partial migration.} Rather than blindly offloading the page migration process to the OS, \verb|move_pages2| enables the MMDB (user application) to control the page migration process via two knobs, namely, \verb|migration_mode| and \verb|nr_max_batched_migration| (cf. Section~\ref{sec:new_sys_call}). The \verb|migration_mode| knob lets the MMDB choose the migration policy from the following options: \verb|MIGRATE_ASYNC|, \verb|MIGRATE_SYNC_LIGHT|, \verb|MIGRATE_SYNC|, and \verb|MIGRATE_SYNC_NO_COPY|.  Section~\ref{sec:new_sys_call} provides a  detailed discussion of
these migration modes. The \verb|nr_max_batched_migration| 
knob lets the MMDB choose the maximum number of pages that can be batched together for migration. {In addition, \texttt{move\_pages2} follows an optimistic approach by allowing partial migration}. We try to make the best use of each system call by migrating as many pages as possible without aborting mid-way. As the invocation of the system call already crosses the costly privilege boundary, it is  natural to amortize the cost by migrating the maximum number of pages rather than aborting the migration midway due to an error.

By default, Linux imposes a strict \verb|MIGRATE_SYNC| migration policy, and sets \verb|nr_max_batched_migration| to $512$, without accounting for the database workload and its migration characteristics. The new system call \verb|move_pages2| that we propose transfers some of the control of the page migration to the MMDB as it has better insight
into the workload than the OS. Linux's page migration policy follows 
an ``abort-on-failure" strategy, i.e., if a page migration fails, the subsequent pages in the list are skipped despite being perfectly eligible for migration. This is detrimental to the performance of the MMDB. The reason is that in a multi-threaded environment, the probability of a page being locked, written back, or being temporarily unavailable is far higher. This not only reduces the throughput of page migration but severely impacts the DBMS performance, as DBMS workers have to give up valuable resources to the kernel threads. Hence, the new system call \verb|move_pages2| bypasses any errors encountered during page migration, records the error, and proceeds with the next page. This improves both page migration and MMDB query throughput (cf. Section~\ref{sec:exp}). 

The rest of this paper proceeds as follows. {Section~\ref{sec:use_case} showcases the use cases of \texttt{move\_pages} in a Main-Memory Database in the context of modern hardware developments}. Section~\ref{sec:sys_call} discusses the \texttt{move\_pages} system call of Linux  and its performance implications. Sections~\ref{sec:new_sys_call} and~\ref{sec:exp} present our proposed custom system call \verb|move_pages2|, and evaluate its performance.  Experiments on a YCSB-like workload shows that compared to \texttt{move\_pages}, \texttt{move\_pages2} improves the performance of a main-memory B$^+$ tree by 2.3$\times$ and 2.6$\times$ in terms of query and page migration throughputs, respectively. 

\section{
MMDB Use Cases of Page Migration 
}
\label{sec:use_case}

\begin{figure*}[t]
    \centering 
    \includegraphics[width=\textwidth]{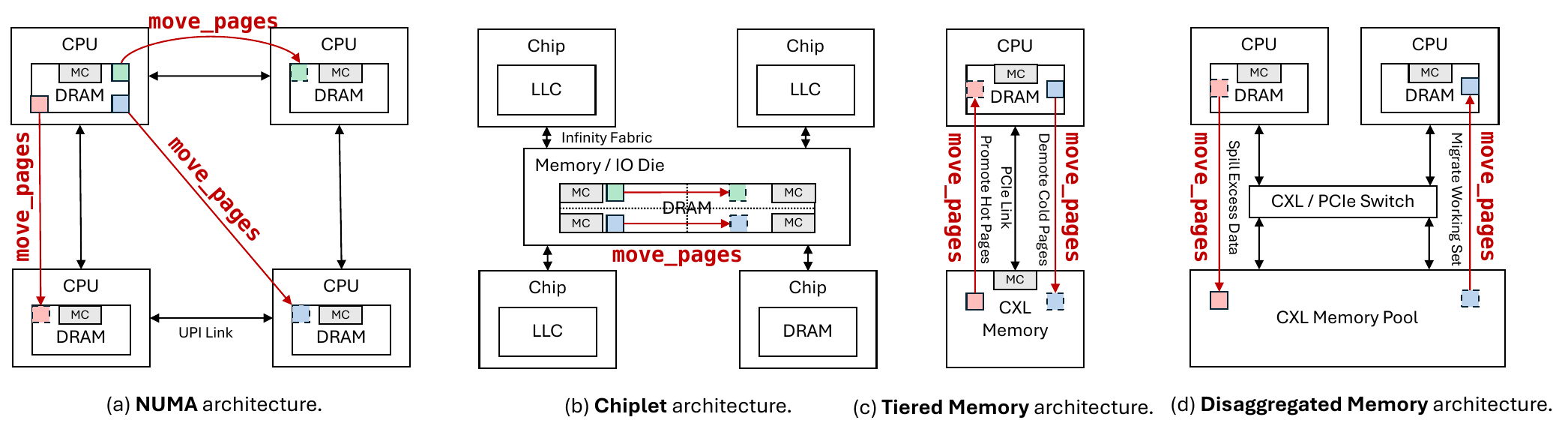}
    \caption{Use-cases of on demand page migration in a Main-Memory Database.}
    \label{fig:use-case}
\end{figure*}

{
We discuss several use cases of the page migration system call, i.e., \texttt{move\_pages}, across four different modern hardware architectures, 
namely,
Non-Uniform Memory Access architecture (NUMA, for short), Chiplet architecture, CXL-enabled Tiered Memory architecture, and CXL-enabled Disaggregated Memory architecture, in the context of a Main Memory Database (MMDB). Figure~\ref{fig:use-case} shows these 4  hardware architectures and illustrates how a Main-Memory Database (MMDB) can leverage the \texttt{move\_pages} system call in each of these environments, where MC refers to the memory controller. We explain each of these use cases below.
}

\vspace{7pt}
\noindent\textbf{1. NUMA Architecture:} {Over the past decade, Non-Uniform Memory Access (NUMA) servers have dominated the hardware landscape, with all CPU vendors, e.g., AMD, Intel, and NVIDIA, adopting NUMA in their processor architectures. Figure~\ref{fig:use-case}a shows a NUMA server with 4 NUMA sockets connected via Ultra-Path Interconnects (UPI, for short). Each NUMA socket comprises a number of compute cores, and a DRAM, referred to as local memory. A socket can access memory of another socket, referred to as remote memory, in a cache-coherent manner, at the expense of additional latency and reduced bandwidth due to  network communication. In a NUMA server, ideally data should be placed as close as possible to the compute core accessing it. This reduces inter-socket communication and thus avoids the additional penalty of accessing remote memory. Recent works in MMDB literature, e.g.,~\cite{PsaroudakisSMSA15,PsaroudakisSMSA16,BangOMPB20} also validate this principle by demonstrating that an adaptive data placement technique across NUMA sockets can improve MMDB performance by up to $4\times$. To achieve this performance improvement, the MMDB dynamically migrates tables across NUMA sockets in order to balance socket utilization and optimize data locality~\cite{PsaroudakisSMSA15,PsaroudakisSMSA16}. The Operating System exposes these NUMA sockets to the MMDB as distinct NUMA nodes, and provides the \texttt{move\_pages} system call to migrate data across sockets. Thus, \texttt{move\_pages} lies on the critical path of any MMDB deployed on NUMA servers.
}

\vspace{7pt}
\noindent\textbf{2. Chiplet Architecture: }{In a chiplet architecture, a processor is composed of multiple smaller chips, termed chiplets. This is in contrast to the traditional monolithic design, where the processor comprises a single unified chip. Each chiplet comprises a set of compute cores and cache slices. They also have their own memory controllers either 
integrated on the same chip (Intel Sapphire Rapids) or located on a separate chip (AMD EPYC Milan). 
This introduces variable access latency similar to NUMA architectures, but within the same 
NUMA socket. For example, accessing memory local to a chiplet can be $4\times$ faster than accessing memory attached to a different chiplet. Thus, the same data access principle for NUMA servers applies to chiplet architectures as well. Figure~\ref{fig:use-case}b shows an AMD EPYC Milan chiplet processor consisting of 5 chiplets. Out of the 5 chiplets, one chiplet termed as the Memory/IO Die houses the unified memory controller and PCIe interfaces. The remaining chiplets are compute chiplets, each containing its own compute cores and cache slices. The chiplets are interconnected 
via high-bandwidth infinity fabric. The Operating System exposes these chiplets to an MMDB as separate NUMA nodes, making \texttt{move\_pages} the default system call for moving data across different chiplets. Only a few MMDBs work over chiplet architectures, e.g.,~\cite{FogliZPBG24,FogliPG24} showcase how database systems can benefit from a chiplet architecture. As the CPU vendors increasingly adopt the chiplet architecture, the \texttt{move\_pages} remains one of the critical system calls for MMDBs to efficiently migrate data across chiplets. 
}

\vspace{7pt}
\noindent\textbf{3. Tiered Memory Architecture.}{ In a tiered memory architecture, a compute chip is connected to multiple types of memory with varying access latency, bandwidth, and capacity. Emerging networking technologies, e.g., Compute Express Link (CXL), have made tiered memory systems a reality. CXL is an interconnect technology that gives the illusion of an expanded memory through a PCIe connected peripheral device in a cache-coherent fashion. 
Refer to the tiered memory architecture in Figure\ref{fig:use-case}c. A NUMA socket is connected to 2 different types of memory, i.e., DRAM memory and CXL-attached memory. While DRAM memory is connected to the socket via high-speed DDR links, the CXL memory is accessed via PCIe bus, resulting in higher access latency and lower bandwidth compared to DRAM. Recent MMDB works, e.g.,~\cite{HaoZYS24,RiekenbrauckWLR24,Guo024} advocate for a tiered buffer pool for CXL-enabled tiered memory architectures, where the hottest pages are placed in the fastest, i.e., DRAM memory, and the coldest pages are placed in the slowest, i.e., CXL memory. The OS exposes CXL memory to an MMDB as a NUMA node without any compute cores~\cite{SunYYKSHJALJ0AX23, abs-2405-14209}. Thus, in a tiered buffer pool, whenever a hot page needs to be promoted to DRAM, or a cold page needs to be evicted to the CXL memory, the \texttt{move\_pages} system call is invoked to perform the migration procedure.
}

\vspace{7pt}
\noindent\textbf{4. Disaggregated Memory Architecture. }{
In a CXL-enabled disaggregated memory architecture, there exist multiple compute nodes and a remote memory pool connected via a switched interconnect, e.g., CXL/PCIe switch~\cite{abs-2503-18140}. The compute nodes may include a small amount of memory while the memory nodes may have limited computational capability. Figure~\ref{fig:use-case}d illustrates a disaggregated memory architecture with 2 compute nodes, and a remote memory pool connected via a CXL switch. Over the past decade, the number of cores per socket has increased. However, the memory capacity of a socket has remained almost the same, effectively reducing the per-core memory bandwidth of modern processors. Moreover, memory prices have plateaued and it still remains one of the most expensive hardware resources in cloud data centers. To mitigate this emerging new ``memory wall"~\cite{memorycentric25}, MMDB communities are increasingly advocating 
for
CXL-centric disaggregated architectures, e.g.,~\cite{memorycentric25,pasha25,Guo024,cxl_switch25}. The effectiveness of these MMDBs on CXL-centric disaggregated architectures heavily relies on the efficiency of page migration, as the core operation of these MMDBs is to seamlessly move data across compute nodes and the remote memory pools. Consider the conventional disk-based external merge sort operation. In the sorting phase, the sort operator fetches small chunks of data from disk to main memory, sorts them, and writes them back to disk. Next, in the merge phase, the operator combines the sorted runs into larger sorted runs. At any point in time, the operator can spill the intermediate data to disk when  main memory is exhausted. In the  disaggregated memory architecture, this translates to the external merge sort operator regularly spilling and fetching data to remote memory pool in place of a disk. 
}


\section{Native Page Migration in Linux}
\label{sec:sys_call}
Linux handles page migration through the system call \verb|move_pages|~\cite{linux_move_pages}. It has been introduced in Linux 2.6.18 by Andi Kleen. Over the years, it has gone through multiple updates in different patches, e.g., 2.6.29, 4.13, 4.17. 

\vspace{7pt}
\noindent\textbf{Signature of \texttt{move\_pages}}: 
The \verb|move_pages| system call accepts six arguments as outlined in its signature: 

\begin{mycodebox}[sys_call_signature]
\begin{small}
\raggedright
\begin{Verbatim}[commandchars=\\\{\}]
long \cpurple{move_pages} (int \cteal{pid}, unsigned long \cteal{count}, void 
    \cteal{*pages[.count]}, const int \cteal{nodes[.count]}, 
    int \cteal{status[.count]}, int \cteal{flags});
\end{Verbatim}
\vspace{0.05mm}
\begin{itemize}
    \item \vteal{pid}. The process ID whose pages are affected.
    \item \vteal{count}. The number of pages to process.
    \item \vteal{pages}. An array of page-pointers that need to be processed.
    \item \vteal{nodes}. An array of target NUMA node IDs.
    \item \vteal{status}. An array to store the migration status of each page.
    \item \vteal{flags}. Specifies the type of pages that need to be processed.
\end{itemize}
\end{small}
\end{mycodebox}

\begin{figure}[bhp]
    \setlength{\belowcaptionskip}{-3mm}
    \centering
    \includegraphics[width=0.97\columnwidth]{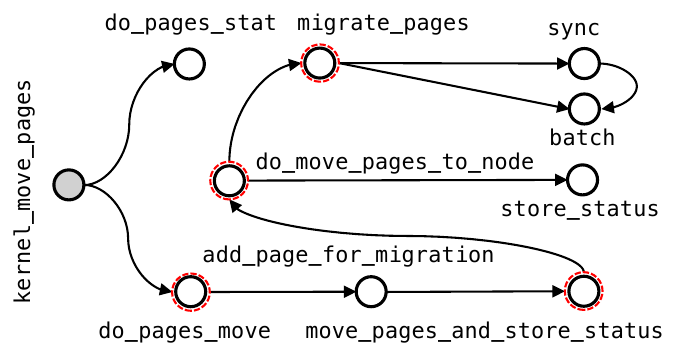}
    \caption{Function call graph of the \texttt{move\_pages} system call. The red circles indicate kernel functions that have been customized to support \texttt{move\_pages2}.}
    \label{fig:inside_calls}
\end{figure}

\vspace{7pt}
\noindent \textbf{Implementation details:} Figure~\ref{fig:inside_calls} presents the function call diagram of  \verb|move_pages|. If the passed \vteal{nodes} list is set to \verb|NULL|, it invokes \verb|do_pages_stat| for querying the location of the input \vteal{pages}. Else, it invokes \verb|do_pages_move| to start migrating the pages in the \vteal{pages} list. The rest of the section discusses the key kernel functions invoked during the execution of the  \verb|move_pages| system call.

\vspace{7pt}
\noindent
\textbf{1. }\vtextbf{do\_pages\_move()}. This kernel function is the primary entry point for migrating pages in the \vteal{pages} list to their respective target destinations. Listing~\ref{lst:do_pages_move} 
provides the steps 
for \verb|do_pages_move|. At a high level, this function handles the preparation phase of the input pages before relocating them. 

\vspace{7pt}
\noindent
1.1. \underline{Disabling LRU Caching:} Before compiling the list of pages to migrate, the kernel disables the Least Recently Used  caching mechanism (LRU) so that no new pages can be added to the LRU list. In Linux, the LRU cache tracks the working memory page set of the application, and evicts pages  when memory is low (Line~\ref{sys:disable_lru_base}). 

\vspace{7pt}
\noindent
1.2. \underline{Migration Round:} Next, the kernel processes each page in the \vteal{pages} list sequentially, and prepares them for migration (Lines ~\ref{sys:loop_start_base} to~\ref{sys:loop_end_base}). The full migration process occurs over multiple rounds, ranging from 1 to \verb|nr_pages|. In each round, the kernel groups consecutive page addresses from the \vteal{pages} list that share the same target NUMA node. It batches 
these
page addresses together, and enqueues them into the \verb|pagelist| kernel queue (Line~\ref{sys:mig_queue_base}). A migration round concludes when the target node changes (Line~\ref{sys:mig_end2_base}) or the kernel reaches the end of the \vteal{page} list (Line~\ref{sys:loop_end_base}). \verb|start| and \verb|i| track the page address boundaries during each migration round. 

\vspace{3pt}
\noindent 
\textbf{Implication on Performance.} The number of migration rounds, i.e., the number of pages in a batch (\verb|pagelist| queue) has implications on \verb|move_pages'| performance as a lower number of migration rounds, i.e., a larger batch size improves the TLB invalidation overhead. The unmapping phase of the \verb|migrate_pages_batch| kernel function provides more detail on this issue. A lower number of migration rounds also reduces the number of calls to the kernel helper function \verb|store_status| that updates the \vteal{status} list. 

\vspace{7pt}
\noindent
1.3. \underline{Handling Errors:} A critical aspect of this kernel function is its error handling that has serious performance implications. If the kernel happens to stumble across any user-space access failures, invalid target nodes, or permission issues for a particular page (Lines~\ref{sys:error1_start_base},~\ref{sys:error1_end_base}), it immediately aborts. Before aborting, it migrates the pages accumulated in the latest migration round (Line~\ref{sys:out_flush_base}) and re-enables the LRU caching mechanism (Line~\ref{sys:enable_lru_base}). Similarly, if any error stems while migrating the pages from the latest round, the kernel re-enables the LRU cache mechanism, and aborts immediately. 

\vspace{3pt}
\noindent
\textbf{Implication on Performance. } Due to the ``abort-on-failure" strategy, all the remaining pages in the input \vteal{pages} list are skipped, even if they are eligible for migration. This becomes more severe in a multi-threaded environment, where multiple threads are working on the same set of pages. 
The reason is 
that the kernel is more likely to encounter 
these 
errors as concurrent thread activity enhances the chances of pages being locked, under writeback, or 
temporarily unavailable. This greatly hampers the overall effectiveness of the migration process, and wastes useful resources. 
\begin{lstlisting}[
    caption={Code snippet of \texttt{do\_pages\_move}. 
    % in \texttt{move\_pages} kernel function.
    }, 
    label={lst:do_pages_move}
    ]
static int do_pages_move(struct mm_struct *mm, nodemask_t task_nodes, unsigned long nr_pages, const void __user * __user *pages, const int __user *nodes, int __user *status,  int flags)
{
  int current_node = NUMA_NO_NODE;
  LIST_HEAD(pagelist);
  int start, i; 
  int err = 0, err1;
  disable_LRU_cache(); (*|\label{sys:disable_lru_base}|*)
  for (i = start = 0; i < nr_pages; i++) { (*|\label{sys:loop_start_base}|*)
    const void __user *p;   // Page to migrate
    int node;               // Destination node
    // ABORT! 
    if (error in copying pages[i] or nodes[i]) (*|\label{sys:error1_start_base}|*)
      goto out_flush;
    // ABORT!
    if (error in handling target device) (*|\label{sys:error1_end_base}|*)
      goto out_flush; 
    // Start a new migration round 
    if (current_node == NUMA_NO_NODE) { (*|\label{sys:mig_start_base}|*)
      current_node = node;
      start = i;
    } 
    // End the current migration round and 
    // migrate the pages collected in the current round
    else if (node != current_node) { (*|\label{sys:mig_end2_base}|*)
      err = move_pages_and_store_status(current_node, 
              &pagelist, status, start, i, nr_pages);
      if (err) // Abort!
       goto out;
      start = i;
      current_node = node;
    }
    // Queue page p for the current migration round
    err = add_page_for_migration(mm, p, current_node, 
            &pagelist, flags & MPOL_MF_MOVE_ALL);(*|\label{sys:mig_queue_base}|*)
    // Page p successfully queued
    if (err > 0) 
      continue; (*|\label{sys:mig_end1_base}|*)
    err = store_status(status, i, err ? : 
          current_node, 1);
    // ABORT!
    if (err) 
      goto out_flush;
    // Migrate the pages collected in the current round
    err = move_pages_and_store_status(current_node, 
            &pagelist, status, start, i, nr_pages);
    if (err) /* Abort! */
      goto out; 
    // Prepare to start a new migration round
    current_node = NUMA_NO_NODE;
  } (*|\label{sys:loop_end_base}|*)
out_flush: (*|\label{sys:out_flush_base}|*)
  // Migrate the pages collected in the 
  // current migration round
  err1 = move_pages_and_store_status(current_node, &pagelist, status, start, i, nr_pages);
  if (err >= 0)
    err = err1;
out: 
  enable_LRU_cache() (*|\label{sys:enable_lru_base}|*)
  return err
}
\end{lstlisting}

\noindent\textbf{2. }\vtextbf{add\_page\_for\_migration().} This helper kernel function is responsible for enqueueing a page into the \verb|pagelist| kernel queue for the next step in the page migration workflow. The process involves mapping the page's virtual address to the page's physical address. Once the page's physical address is mapped, the kernel isolates the page from the LRU cache 
(\verb|folio_isolate_lru|). Finally, when  isolation succeeds, it puts the page into the \verb|pagelist| queue. 

\vspace{7pt}
\noindent\textbf{3. }\vtextbf{move\_pages\_and\_store\_status().} This helper kernel function (cf. Listing~\ref{lst:move_pages_store_status}) invokes the \verb|do_move_pages_to_node| kernel function to relocate the pages accumulated during a migration round. Once  relocation is complete, it invokes the \verb|store_status| kernel function to record the migration status of all the pages in the latest migration round.

\begin{lstlisting}[
    style=mystyle, 
    caption={Code snippet of \texttt{move\_pages\_and\_store\_status}. 
    % in \texttt{move\_pages}.
    },  label={lst:move_pages_store_status}
]
static int move_pages_and_store_status(int node, struct list_head *pagelist, int __user *status, int start, int i, unsigned long nr_pages)
{
  int err;
  if (list_empty(pagelist))
    return 0;
  err = do_move_pages_to_node(pagelist, node);
  if (err) {
    if (err > 0)
      err += nr_pages - i;
    return err;
  }
  return store_status(status, start, node, i - start);
}
\end{lstlisting}

\vspace{7pt}
\noindent\textbf{4. }\vtextbf{do\_move\_pages\_to\_node().} This helper function (cf. Listing~\ref{lst:do_move_pages_to_node}) sets the target destination NUMA node for the pages in the \verb|pagelist| queue, and invokes the \verb|migrate_pages| kernel function. If the migration is unsuccessful, i.e., the kernel has failed to migrate some or all the pages, it invokes the \verb|putback_movable_pages| kernel function on the non-migrated pages in the \verb|pagelist| queue. This ensures that the non-migrated pages from the latest migration round are reintegrated into the kernel's memory management structure. Recall that, during the \verb|add_pages_to_migrate| system call, all the pages in the \verb|pagelist| 
are
isolated from the LRU cache. The \verb|putback_movable_pages| just reverts the non-migrated pages back to their original state. 

\begin{lstlisting}[
    style=mystyle, 
    caption={Code snippet of and \texttt{do\_move\_pages\_to\_node}.
    % in \texttt{move\_pages}.
    },  label={lst:do_move_pages_to_node}
]
static int do_move_pages_to_node(struct list_head *pagelist, int node)
{
  int err;
  // Set the target node for the current 
  // migration round
  struct migration_target_control mtc = {  
    .nid = node,
    .gfp_mask = GFP_HIGHUSER_MOVABLE | __GFP_THISNODE,
  };
  err = migrate_pages(pagelist, alloc_migration_target, NULL, (unsigned long)&mtc, MIGRATE_SYNC, MR_SYSCALL, NULL);
  // Restore the original state of the non-migrated 
  // pages
  if (err)
    putback_movable_pages(pagelist);  
  return err;
}
\end{lstlisting}
\vspace{7pt}
\noindent\textbf{5. }\vtextbf{migrate\_pages} This kernel function is responsible for migrating all the pages in a migration round accumulated in the \verb|pagelist| kernel queue. At a high level, \verb|migrate_pages| sets the migration policy, and processes the pages in the \verb|pagelist| queue by dividing them into multiple sub-groups. 

\vspace{7pt}
\noindent
5.1. \underline{Segmenting \texttt{pagelist}:} Recall that \verb|pagelist| contains the list of page addresses in the latest migration round that require migration. The \verb|migrate_pages| kernel call divides the page addresses in the \verb|pagelist| queue into multiple sub-groups of size \verb|NR_MAX_BATCHED_MIGRATION=512|, i.e., each sub-group can contain at most \verb|NR_MAX_BATCHED_MIGRATION| 
page addresses. 

\vspace{7pt}
\noindent
5.2. \underline{Invoking \texttt{migrate\_pages\_sync}:} For each sub-group, the  kernel strictly invokes the \verb|migrate_pages_sync| kernel function. 


\vspace{7pt}
\noindent\textbf{6. }\vtextbf{migrate\_pages\_sync().} Despite the name, the synchronous approach initially employs an asynchronous phase for page migration, and falls back to a synchronous phase if any pages remain. During both phases, the kernel 
maintains
statistics on successes, failures, and splits throughout the migration round in a \verb|migrate_pages_stats| struct.

\vspace{7pt}
\noindent
6.1. \underline{Asynchronous Phase: } Initially, the kernel undertakes an optimistic approach, and calls the \verb|migrate_pages_batch| function in \verb|MIGRATE_ASYNC| mode to migrate all the pages in the input sub-queue. It attempts this operation up to 3$\times$, following the default Linux setting \verb|NR_MAX_MIGRATE_ASYNC_RETRY=3|.

\vspace{3pt}
\noindent
\textbf{Implication on Performance.}  In the asynchronous phase, the kernel employs the \verb|MIGRATE_ASYNC| mode that is faster than the \verb|MIGRATE_SYNC| mode as it handles pages in batches. Moreover, in case of failures, the number of retries 
is
amortized over at most \verb|NR_MAX_BATCHED_MIGRATION| 
pages. 

\vspace{7pt}
\noindent
6.2. \underline{Synchronous Phase:} Once the asynchronous attempt finishes, the kernel accumulates all the non-migrated pages from the asynchronous phase and reverts to a more pessimistic approach. For each non-migrated page, the kernel invokes the \verb|migrate_pages_batch| function in \verb|MIGRATE_SYNC| mode. If the migration fails, the kernel retries up to $7\times$ following the default Linux setting \verb|NR_MAX_MIGRATE_SYNC_RETRY=7|. Notice that, in the synchronous phase, the kernel invokes  \verb|migrate_pages_batch| with a single page, whereas in the asynchronous phase, it invokes the same kernel function with 
all the pages in the sub-queue. 

\vspace{3pt}
\noindent
\textbf{Implication on Performance.} In the synchronous phase, the kernel employs the \verb|MIGRATE_SYNC| mode that is much slower as each page is processed individually. 
Moreover, 
in case of failures, the synchronous phase can 
retry up to 
$7\times$ for each of the \verb|NR_MAX_BATCHED_MIGRATION| pages.

\vspace{7pt}
\noindent\textbf{7. }\vtextbf{migrate\_pages\_batch().} This kernel function facilitates the migration of a single page or a batch of pages. In 
the
\verb|MIGRATE_SYNC| mode, \verb|migrate_pages_batch| migrates a single page, and in \verb|MIGRATE_ASYNC| mode, it migrates a batch of pages. 

\vspace{7pt}
\noindent
7.1. \underline{Unmapping Phase:} The kernel processes each page in the input list 
one at a time, 
and attempts to unmap the page so that its data can be safely copied to a new page. It locks each page in the list, and completes the writeback of the page. 
Also, the kernel
allocates a new page, say $p_n$, for the old page, say $p_o$, to move to, and locks 
$p_n$
immediately. If the page $p_o$ is mapped, the kernel unmaps it by invoking \verb|try_to_migrate|. This invalidates all the page table references to 
$p_o$
by flagging it as a migration page table entry (PTE), and issues a TLB shootdown. During the \verb|MIGRATE_PAGES_ASYNC| mode, the TLB shootdown is deferred until all the pages in the list 
are
processed. The kernel makes several passes over the pages in the list to unmap as many pages as possible for the next phase~\cite{page_migration}.

\vspace{3pt}
\noindent
\textbf{Implication on Performance.} The TLB shootdown is an expensive step as it involves sending inter-processor interrupts (IPIs) to all relevant CPUs to invalidate specific TLB entries. In the \verb|MIGRATE_ASYNC| mode, the kernel can amortize the cost of a single TLB shootdown over at best \verb|NR_MAX_BATCHED_MIGRATION| pages. On the other hand, in the \verb|MIGRATE_SYNC| mode, the kernel may 
have to issue \verb|NR_MAX_BATCHED_MIGRATION| TLB shootdowns, 
significantly degrading 
the migration performance. 

\vspace{7pt}
\noindent
7.2. \underline{Moving Phase:} As in the unmapping phase, the kernel takes multiple passes over the unmapped pages from the unmapping phase, and processes each page sequentially. It calls \verb|migrate_folio_move| to  move the page to its target destination. Once migration succeeds, all page locks both old and new are dropped. Finally, the new page is moved to the LRU list. 

\vspace{7pt}
\noindent
7.3. \underline{Cleanup Phase:} During both 
the 
unmapping and moving 
phases,
if a page 
cannot
be unmapped or migrated, it is reverted 
to its original state. The kernel 
keeps track of detailed 
statistics 
on successes and failures in 
the 
\verb|migrate_pages_stats| struct.

\vspace{7pt}
\noindent\textbf{8. }\vtextbf{do\_pages\_stat(). }
This kernel function is the primary entry point for querying the location of each page in the input \vteal{pages} list. The kernel copies the virtual page addresses from the \vteal{pages} list in batches of 16 pages,
and then processes each page separately. First, it validates the input virtual page address by checking if it indeed falls within the virtual memory area of the calling process through a \verb|vma_lookup| call. Once the page address validation  is confirmed, 
the kernel 
scans
the page table (\verb|follow_page|) to retrieve the physical address mapped to the corresponding 
virtual 
address. Finally, the kernel invokes \verb|page_to_nid| to identify the NUMA ID of the corresponding page and stores 
the NUMA ID 
in the \vteal{status} array.

\section{The  New Custom System Call: \vtextbf{move\_pages2 
}}
\label{sec:new_sys_call}
For the new proposed custom implementation of the page migration system call, we exclude the 2 parameters \textcolor{teal}{\texttt{pid}} and \textcolor{teal}{\texttt{flags}}, and instead include parameters, 
\textcolor{teal}{\texttt{migrate\_mode}} and \textcolor{teal}{\texttt{nr\_max\_batched\_migration}} to facilitate more fine-grained control over the page migration mechanism. Our custom implementation 
sets by default the \vteal{pid}\verb|=0|, i.e., the calling process can 
migrate only those pages that belong to 
itself. We set the default \vteal{flags}\verb|=MPOL_MF_MOVE|, i.e., only pages that exclusively belong to the calling process can be migrated. 

\vspace{7pt}
\noindent\textbf{Signature of \texttt{move\_pages2}}: The signature of the proposed custom implementation of \verb|move_pages2| is as follows. 

\begin{mycodebox}
\begin{small}
\raggedright
\begin{Verbatim}[commandchars=\\\{\}]
long \cpurple{move_pages2} (unsigned long \cteal{count}, void \cteal{*pages[.count]}, 
    const int \cteal{nodes[.count]}, int \cteal{status[.count]}, 
    enum migrate_mode \cteal{mode}, int \cteal{nr_max_batched_migration});
\end{Verbatim}
\vspace{0.05mm}
\begin{itemize}
    \item \textcolor{teal}{\texttt{count}}. The number of pages to process.
    \item \textcolor{teal}{\texttt{pages}}. An array of page-pointers that need to be processed.
    \item \textcolor{teal}{\texttt{nodes}}. An array of target NUMA node IDs.
    \item \textcolor{teal}{\texttt{status}}. An array to store the migration status of each page.
    \item \textcolor{teal}{\texttt{mode}}. The mode of migration. 
    \item \textcolor{teal}{\texttt{nr\_max\_batched\_migration}}. The maximum number of pages that can be accumulated before TLB shootdown is invoked.
\end{itemize}
\end{small}
\end{mycodebox}

\vspace{7pt}
\noindent \textbf{Implementation details:} We introduce approximately 150 lines of code (LOC) changes across 
the following four kernel functions: 
\verb|do_pages_move|, \verb|move_pages_and_store_status|, \verb|do_move_pages_to_node|, and \verb|migrate_pages| to implement the proposed custom system call \verb|move_pages2| (cf. Figure~\ref{fig:inside_calls}). We implement \verb|move_pages2| in Linux kernel 6.8.0 on Ubuntu 24.04 LTS. 
{The Linux kernel disk image is available on CloudLab~\cite{DuplyakinRMWDES19} and can be accessed here\footnote{\href{}{urn:publicid:IDN+utah.cloudlab.us+image+pmoss-PG0:LINUX6.8.12\_MIGRATE}}. The rest of the section discusses the two configuration knobs that the \texttt{move\_pages2} system call exposes to the user application, e.g., an MMDB, followed by the modifications introduced to each of the aforementioned kernel functions.}

\vspace{7pt}
\noindent\textbf{1. Page Migration Mode.} The Linux kernel supports 
four page migration modes. 
\verb|move_pages2| includes \vteal{mode} as one of the parameters so that the MMDB can choose the migration mode 
according to its specific use-case.
\begin{itemize}[leftmargin=*]
    \item \verb|MIGRATE_ASYNC|. The kernel performs asynchronous migration, i.e., it proceeds with the next page even if the current page migration fails, making it a non-blocking approach.
    \item \verb|MIGRATE_SYNC|. The kernel performs synchronous migration, i.e., 
    the kernel
    blocks until the current page migration succeeds.
    \item \verb|MIGRATE_SYNC_LIGHT|. Unlike \verb|MIGRATE_SYNC|, this  approach 
    does not block on page writebacks to reduce the stall time. 
    \item \verb|MIGRATE_SYNC_NO_COPY|. Unlike \verb|MIGRATE_SYNC|, this blocking approach skips copying the old page data to the new page. It only copies metadata, and is particularly helpful when moving HugeTLB pages, 
    where data transfer is unnecessary.
\end{itemize}

\vspace{7pt}
\noindent\textbf{2. Maximum Batch Size.} \verb|move_pages2| includes \vteal{nr\_max\_batched}
\vteal{\_migration} 
as one of the parameters, as it regulates the number of pages that can be simultaneously migrated in 
the \verb|MIGRATE_ASYNC| mode. Recall that \vteal{nr\_max\_batched\_migration} impacts the performance of the page migration process by amortizing the cost of TLB shootdown over \vteal{nr\_max\_batched\_migration} number of memory pages. 

\vspace{7pt}
\noindent\textbf{3. }\vtextbf{do\_pages\_move2().} Listing~\ref{lst:cus_do_pages_move} summarizes the steps 
of the custom kernel function \verb|do_pages_move2|. The highlighted code illustrates the modifications over the original 
\verb|do_pages_move| kernel function. Among the 3 steps, i.e., disable LRU caching, perform the migration round, and handle errors, we modify the error-handling step to ensure 
partial page migration, i.e., page migration continues despite 
errors. 

\vspace{7pt}
\noindent3.1. \underline{Error Handling:} If the kernel happens to 
encounter
any user-space access failures, invalid target node, or permission issues for a particular page, rather than immediately aborting, it records the error in the \vteal{status} array. Next, it invokes the custom kernel function \verb|move_pages_store_status2| to initiate the migration process of the pages accumulated in the latest migration round. Once the migration is complete, the kernel initiates a new migration round, and proceeds with processing the next page from the \vteal{pages} list (Lines~\ref{new_sys:handle_error_start} to~\ref{new_sys:handle_error_end}). Similarly, if the kernel encounters any error during the migration process itself, it records the error in the \vteal{status} array, migrates the pages in the latest round, and continues with the migration process (Lines~\ref{new_sys:migrate_error_start} to~\ref{new_sys:migrate_error_end}).

\begin{lstlisting}[
    style=mystyle, 
    caption={Code snippet of \texttt{do\_pages\_move2}.
    % in \texttt{move\_pages2}. 
    The modifications introduced over the original implementation are highlighted.},  
    label={lst:cus_do_pages_move}
]
static int do_pages_move2(struct mm_struct *mm, nodemask_t task_nodes, unsigned long nr_pages, const void __user * __user *pages, const int __user *nodes, int __user *status,  int flags, @enum migrate_mode mode, int nr_max_batched_migration@)
{
  int current_node = NUMA_NO_NODE;
  LIST_HEAD(pagelist);
  int start, i;
  int err = 0, err1;
  disable_LRU_cache(); (*|\label{sys:disable_lru}|*)
  for (i = start = 0; i < nr_pages; i++) { (*|\label{sys:loop_start}|*)
    const void __user *p;  // Page to migrate
    int node;              // Destination node
    @if (error in copying pages[i] or nodes[i])(*|\label{sys:error1_start}|*)
      goto handle_error;
    if (error in handling target device, i.e., NUMA node)(*|\label{sys:error1_end}|*)
      goto handle_error;@
    if (current_node == NUMA_NO_NODE) { (*|\label{sys:mig_start}|*)
      // Start a new migration round
      current_node = node;
      start = i;
    } else if (node != current_node) { (*|\label{sys:mig_end2}|*)
      // End the current migration round and migrate 
      // the pages collected in the current round
      @err = move_pages_and_store_status2(current_node,
              &pagelist, status, start, i + 1, nr_pages, 
              mode, nr_max_batched_migration);@ 
      @if (err) goto migrate_error;@
      start = i;
      current_node = node;
    }
    // Queue page p for the current migration round 
    err = add_page_for_migration(mm, p, current_node,
            &pagelist, flags & MPOL_MF_MOVE_ALL);(*|\label{sys:mig_queue}|*)
    // Page p successfully queued
    if (err > 0) 
      continue;  (*|\label{sys:mig_end1}|*)
    err = store_status(status, i, err ? : current_node, 1);
    @if (err) 
      goto migrate_error; @
    @migrate_error:(*|\label{new_sys:migrate_error_start}|*)
      if (i + 1 == nr_pages || err < 0) {
        // Migrate pages collected in the current round
        err1 = move_pages_and_store_status2(...);
        if (err >= 0) 
          err = err1;
        // Prepare to start a new migration round 
        current_node = NUMA_NO_NODE; 
      }
      // Continue the migration process
      continue; (*|\label{new_sys:migrate_error_end}|*)
    handle_error:(*|\label{new_sys:handle_error_start}|*)
      err1 = store_status(status, i, err, 1);
      if (err1) 
        err = err1;
      // Migrate pages collected in the current 
      // migration round
      err1 = move_pages_and_store_status2(...);
      if (err >= 0) 
        err = err1; 
      // Prepare to start a new migration round  
      current_node = NUMA_NO_NODE; @ (*|\label{new_sys:handle_error_end}|*)
  }(*|\label{sys:loop_end}|*)
out_flush: (*|\label{sys:out_flush}|*)
  @if (current_node != NUMA_NO_NODE) {
    err1 = move_pages_and_store_status2(current_node, 
            &pagelist, status, start, i, nr_pages, 
            mode, nr_max_batched_migration);
    if (err >= 0) err = err1;@
  }
out: 
  enable_LRU_cache(); (*|\label{sys:enable_lru}|*)
  return err;
}
\end{lstlisting}

\noindent\textbf{4. } \vtextbf{move\_pages\_and\_store\_status2()}. We make minimal changes to the logic of the native \verb|move_pages_and_store_status| kernel function. Our custom implementation extends this function's functionality by adding additional parameters: \vteal{mode}, and \vteal{nr\_max\_batched\_migration}. The signature of the \texttt{move\_pages\_and\_store\_status2} is as follows.


\begin{mycodebox}
\begin{small}
\raggedright
\begin{Verbatim}[commandchars=\\\{\}]
static int \cpurple{move_pages_and_store_status2} (int node, struct 
    list_head *pagelist, int __user *status, int start, 
    int i, unsigned long nr_pages, \cteal{enum migrate_mode mode}, 
    \cteal{int nr_max_batched_migration});
\end{Verbatim}
\end{small}
\end{mycodebox}

\vspace{1pt}
\noindent\textbf{5. }\vtextbf{do\_move\_pages\_to\_node2().}  This custom implementation extends the native kernel function by adding the migration mode and batch size parameters. The signature of the \texttt{do\_move\_pages\_to\_node2} is as follows.


\begin{mycodebox}
\begin{small}
\raggedright
\begin{Verbatim}[commandchars=\\\{\}]
static int \cpurple{do_move_pages_to_node2} (struct list_head 
     *pagelist, int node, \cteal{enum migrate_mode mode}, 
    \cteal{int nr_max_batched_migration});
\end{Verbatim}
\end{small}
\end{mycodebox}

\vspace{7pt}
\noindent\textbf{6. } \vtextbf{migrate\_pages2().} Our custom implementation (cf. Listing~\ref{lst:migrate_pages2}) enables the DBMS to override the default Linux migration mode, i.e., \verb|MIGRATE_SYNC|. For the \verb|MIGRATE_SYNC| mode, the kernel follows the default blocking path (Line~\ref{mig:choose_mode1}). Consequently, for the rest of the migration modes, the kernel follows the non-blocking path (Line~\ref{mig:choose_mode2}). Beyond the default Linux path, the DBMS now has more precise control over the migration process through different variants of the synchronous migration mode, i.e., \verb|MIGRATE_SYNC_LIGHT| and \verb|MIGRATE_SYNC_NO_COPY|, depending on the workload requirements.
We further extend the functionality of the native \verb|migrate_pages| kernel function by letting the DBMS choose the batch size during 
migration 
(Lines~\ref{mig:cut_1},~\ref{mig:cut_2}).

\begin{lstlisting}[
    caption={Code snippet of \texttt{migrate\_pages2}. 
    % in \texttt{move\_pages2}. 
    The modifications introduced over the original implementation are highlighted.}, 
    label={lst:migrate_pages2}
]
int migrate_pages2(struct list_head *from, new_folio_t get_new_folio, free_folio_t put_new_folio, unsigned long private, @enum migrate_mode mode@, int reason, unsigned int *ret_succeeded, int @nr_max_batched_migration@)
{
again:
  nr_pages = 0;
  // Iterate over the pages in "from"
  list_for_each_entry_safe(folio, folio2, from, lru) { 
    nr_pages += folio_nr_pages(folio);
    if (nr_pages >= @nr_max_batched_migration@) break;(*|\label{mig:cut_1}|*)
  }
  // Divide the pages in "from" into smaller 
  // sub-groups, i.e., "folios" 
  if (nr_pages >= @nr_max_batched_migration@)(*|\label{mig:cut_2}|*)
    list_cut_before(&folios, from, &folio2->lru);
  else
    list_splice_init(from, &folios);
  // Migrate the pages in the sub-group, i.e., "folios"
  @if (migration_mode == MIGRATE_SYNC)(*|\label{mig:choose_mode1}|*)
    rc = migrate_pages_sync(&folios, get_new_folio,
          put_new_folio, private, MIGRATE_SYNC, reason,
          &ret_folios,&split_folios, &stats); 
  else(*|\label{mig:choose_mode2}|*)
    rc = migrate_pages_batch(&folios, get_new_folio,
          put_new_folio, private, mode, reason, 
          &ret_folios, &split_folios, &stats, 
          NR_MAX_MIGRATE_PAGES_RETRY);@
  // Store the non-migrated pages in "ret_folios" for potential retries
  list_splice_tail_init(&folios, &ret_folios);  
  if (rc < 0) {
    rc_gather = rc;
    list_splice_tail(&split_folios, &ret_folios);
    goto out;
  }
  rc_gather += rc;
  // Prepare the next "folios"
  if (!list_empty(from)) 
    goto again;
 out:
  // Move back the non-migrated pages in the current 
  // migration round to "folios" for potential 
  // retry in the next migration round
  list_splice(&ret_folios, from); 
  if (list_empty(from)) 
    rc_gather = 0;
  return rc_gather;
}
\end{lstlisting}

\begin{figure*}[htbp]
    \captionsetup{aboveskip=-0.1pt}
    \centering
    \includegraphics[width=\textwidth]{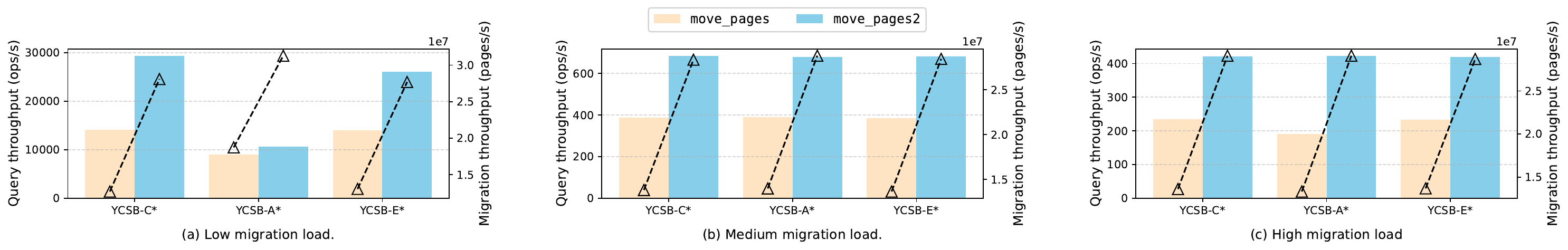}
    \caption{Performance of \texttt{move\_pages2} on NUMA architecture.}
    \label{exp:exp1_intel}
\end{figure*}
\begin{figure*}[htbp]
    \captionsetup{aboveskip=-0.1pt}
    \centering
    \includegraphics[width=\textwidth]{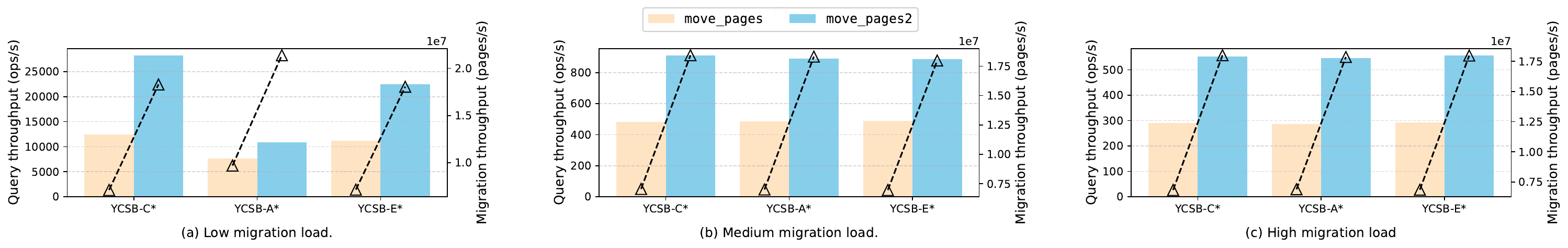}
    \caption{Performance of \texttt{move\_pages2} on Chiplet architecture.}
    \label{exp:exp1_amd}
\end{figure*}

\section{Performance Evaluation}
\label{sec:exp}
Recall the use cases of page migration presented in \S~\ref{sec:use_case} in the context of an MMDB. 
In this section, we focus on the first 2 use cases and evaluate the performance of the two system calls \verb|move_pages| and \verb|move_pages2| in the context of an MMDB on 2 different hardware architectures, i.e., NUMA architecture (cf. \S~\ref{subsec:exp_numa}) and chiplet architecture (cf. \S~\ref{subsec:exp_chiplet}).
We show that our proposed system call has significant performance improvements in both query throughput and page migration throughput. Additionally, we evaluate the impact of migration modes and batch size on the performance of \verb|move_pages2| (cf. \S~\ref{subsec:exp_sen}) ). 

\subsection{Evaluation on NUMA Architecture}
\label{subsec:exp_numa}
\noindent \textbf{Hardware:} {The NUMA hardware is a dual socket 20 ($\times$2) core Intel(R) Xeon(R) Silver 4114 processor, codenamed Intel Skylake, running Ubuntu 24.04.2 LTS on a CloudLab~\cite{DuplyakinRMWDES19} c220g5 node. The respective size of its L1d, L1i, L2, and LLC cache are 640 KB, 640 KB, 20 MB and 27.5 MB, respectively. The machine is equipped with 188GB of DRAM memory. The CPU clock frequency is 2.2 GHz.} 

\vspace{7pt}
\noindent\textbf{Experiment Setting:} {We run a YCSB-like benchmark~\cite{CooperSTRS10} on a main memory B$^+$-Tree~\cite{btreecode}. The B$^+$-Tree implementation follows the optimistic lock coupling (OLC) approach for concurrent operations. Each B$^+$-Tree node is 4 KB, i.e., can hold a maximum of 256 floating point entries. The workload comprises a mixture of the traditional YCSB-A, YCSB-C, YCSB-E workload, and migration queries. Each migration query is responsible for migrating a set of index nodes, i.e., 4 KB memory pages to a designated NUMA node. These migration queries help evaluate the performance of both system calls. To simulate the migration load on the system, we use 3 different workload configurations, i.e., low migration load (\texttt{X/Y} = 99.99/0.01), medium migration load (\texttt{X/Y} = 75/25), and  high migration load (\texttt{X/Y}= 50/50). Both \texttt{X} and \texttt{Y} control the frequency of migration queries, simulating how frequently page migration is handled. For example, 75/25 YCSB-A$^*$ comprises \texttt{X} $=75\%$ of the default YCSB-A workload and \texttt{Y} $=25\%$ of migration queries. Initially, the index is loaded with 300M records, and it grows to 500M records as
more insertions are performed. Each index record has a 64-bit key and a 64-bit value. 
The source code is available here.\footnote{https://github.com/purduedb/PMOSS/tree/test\_move\_pages2}
}

\vspace{7pt}
\noindent\textbf{Overall Performance:} 
{Figure~\ref{exp:exp1_intel} gives the performance of a main-memory B$^+$-Tree under low, medium, and high migration loads. 
Under a low migration load, the B$^+$-Tree implementation using \texttt{move\_pages2} outperforms the B$^+$-Tree implementation using \texttt{move\_pages} by 1.2$\times$, 2.1$\times$ and 1.9$\times$ in terms of B$^+$-Tree query throughput for YCSB-A$^*$, YCSB-C$^*$, and YCSB-E$^*$ workloads, respectively. 
As the migration load increases, the performance gap between the B$^+$-Tree implementation using \texttt{move\_pages2} and the B$^+$-Tree implementation using \texttt{move\_pages} for the YCSB-A$^*$ workload increases. Under medium and high migration loads, the B$^+$-Tree implementation using \texttt{move\_pages2} outperforms the B$^+$-Tree implementation using \texttt{move\_pages} 
by 1.7$\times$ and 2.2$\times$, respectively. 
This is due to the nature of default YCSB-A workload that comprises 50\% writes. This introduces more contention, and increases the likelihood of a memory page being locked. It results in an increased number of ``aborted" \texttt{move\_pages} system call, in contrast to the \texttt{move\_pages2} system call that allows partial migration. The B$^+$-Tree page migration throughput supports this claim, i.e., \texttt{move\_pages} migrates 1.67$\times$, 2.05$\times$, 2.18$\times$ less B$^+$-Tree memory pages than \texttt{move\_pages2} for the YCSB-A$^*$ workload under low, medium and high migration load, respectively. 
For read-dominant workloads, i.e., YCSB-C$^*$, YCSB-E$^*$, as the migration load increases, the performance gap between the 
the B$^+$-Tree implementation using \texttt{move\_pages2} and the B$^+$-Tree implementation using \texttt{move\_pages} 
decreases. Under medium and high migration load, \texttt{move\_pages2} outperforms \texttt{move\_pages} by improving the B$^+$-Tree query throughput by 1.8$\times$ for both read-dominant workloads. 
In an MMDB, page migration demand may arise due to sudden changes in the workload, and/or sudden shifts in the working set of a NUMA socket. A higher page migration throughput ensures the MMDB 
copes
with the migration demand, and maintains the system's optimal state. 
The B$^+$-Tree implementation using \texttt{move\_pages2} also outperforms the B$^+$-Tree implementation using \texttt{move\_pages} in 
this regard. Under low migration load, \texttt{move\_pages2} migrates 1.7$\times$, 2.2$\times$ and 2.1$\times$ more B$^+$-Tree memory pages than \texttt{move\_pages} for YCSB-A$^*$, YCSB-C$^*$, and YCSB-E$^*$ workloads, respectively. As the migration load increases, the performance of the B$^+$-Tree implementation using \texttt{move\_pages2} remains stable, and delivers up to 2.2$\times$ better page migration throughput over the B$^+$-Tree implementation using the native Linux baseline. 
}

\subsection{Evaluation on Chiplet Architecture}
\label{subsec:exp_chiplet}
\noindent\textbf{Hardware:} {The chiplet hardware is an AMD EPYC 7452 processor with 32 cores, codenamed AMD Rome, running Ubuntu 24.04.2 LTS on a CloudLab~\cite{DuplyakinRMWDES19} d6515 node. The machine has 8 chiplets, i.e., Core Complex Dies (CCD, for short), where each chiplet is equipped with 4 cores and a 128 MB L3 cache. Each core has its own L1d, L1i, L2 cache. The respective sizes of the caches are 1 MB, 1 MB and 16 MB, respectively. The machine is equipped with 125GB of DRAM memory. The CPU clock frequency is 2.2 GHz. The BIOS of the machine is configured to use a NUMA Per Socket (NPS, for short) value of 4.}

\vspace{7pt}
\noindent\textbf{Experiment Setting:} {We follow the same experiment settings stated in 
Section~\ref{subsec:exp_numa}.}

\vspace{7pt}
\noindent\textbf{Overall Performance:} {Figure~\ref{exp:exp1_amd} gives the performance of a main-memory B$^+$-Tree under low, medium and high migration load. Under low migration load, the B$^+$-Tree implementation using \texttt{move\_pages2} outperforms the B$^+$-Tree implementation using \texttt{move\_pages} by 1.4$\times$, 2.3$\times$ and 2.0$\times$  in terms of query throughput for YCSB-A$^*$, YCSB-C$^*$, and YCSB-E$^*$ workloads, respectively. 
The gap in performance between the two B$^+$-Tree implementations using \texttt{move\_pages2} and \texttt{move\_pages} system calls is more significant in the chiplet architecture than in the NUMA architecture. This is due to the greater difference in access latency between the chiplets in the AMD Rome architecture compared to the Intel Skylake architecture.
As the migration load increases, the performance gap between the two 
B$^+$-Tree implementations 
increases under write-dominant workloads and decreases under read-dominant workloads. This is 
consistent with the B$^+$-Tree experiments on the NUMA architecture. For write-heavy YCSB-A$^*$ workload, \texttt{move\_pages2} outperforms \texttt{move\_pages} by improving the B$^+$-Tree query throughput 1.9$\times$ (36\% increase in performance gap) under high migration load.
Likewise, for read-heavy YCSB-C$^*$ workload, \texttt{move\_pages2} outperforms \texttt{move\_pages} by improving the B$^+$-Tree query throughput by 1.9$\times$ (17\% decrease in performance gap) under high migration load.
Across all workloads and 
varying 
migration loads, 
the B$^+$-Tree implementation using \texttt{move\_pages2} consistently maintains high page migration throughput by up to 2.6$\times$ than the B$^+$-Tree implementation using \texttt{move\_pages}. 
}

\begin{figure*}[t]
    \captionsetup{belowskip=-6pt}
    \begin{subfigure}{0.49\textwidth}
		\centering
	    \includegraphics[width=\columnwidth]{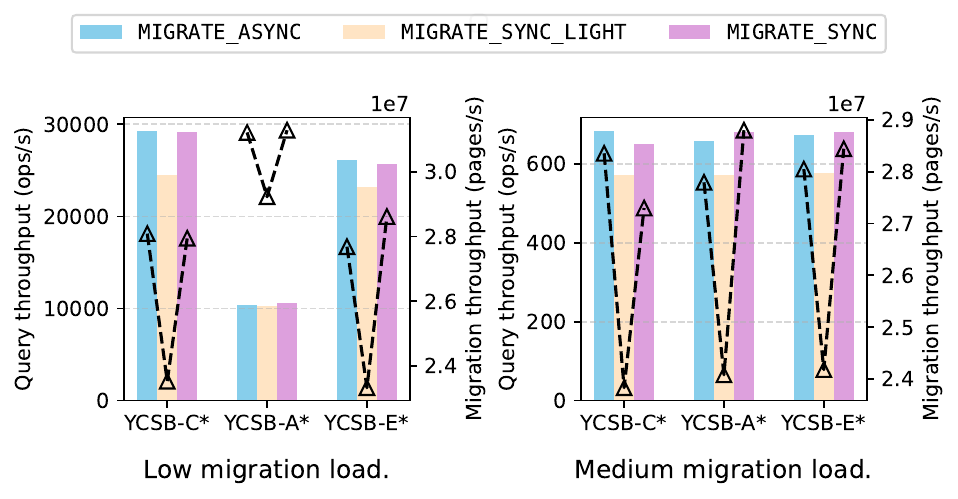}
         \caption{NUMA architecture.}
        \label{exp:exp2_intel}
	\end{subfigure}
	\begin{subfigure}{0.49\textwidth}
		\centering
	    \includegraphics[width=\columnwidth]{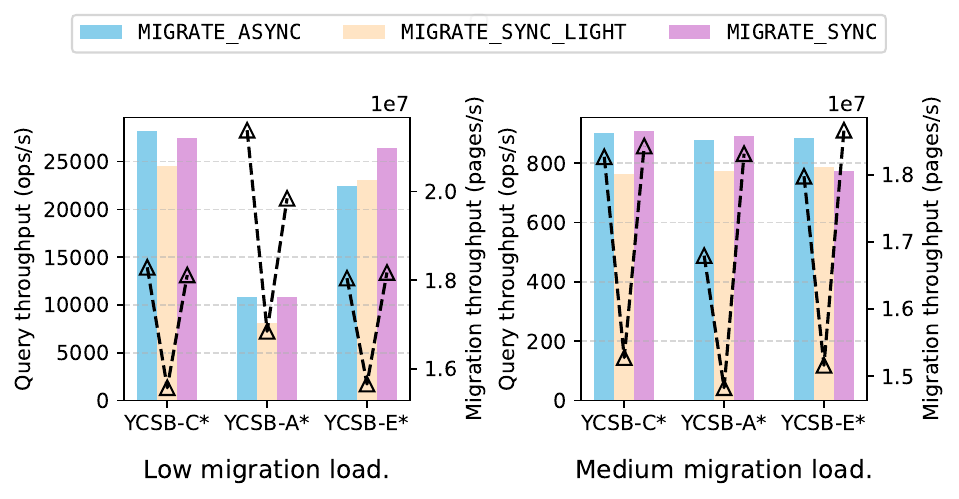}
         \caption{Chiplet architecture.}
        \label{exp:exp2_amd}
	\end{subfigure}
    \caption{Performance of different migration modes in \texttt{move\_pages2}.}
	\label{Fig:exp2}
\end{figure*}

\subsection{Parameter Sensitivity of \texttt{move\_pages2}}
\label{subsec:exp_sen}

\noindent\textbf{Migration Mode:} 
{\texttt{move\_pages2} introduces the \vteal{migration\_mode} knob that lets 
the B$^+$-Tree 
choose the desired migration policy from the following 3 options: \texttt{MIGRATE\_ASYNC}, \texttt{MIGRATE\_SYNC\_LIGHT}, and \texttt{MIGRATE\_SYNC}. Notice that despite using the built in migration modes, e.g., \texttt{MIGRATE\_SYNC} of the Linux kernel, \texttt{move\_pages2} allows partial migration through the custom \texttt{do\_pages\_move2} kernel function (cf. Listing~\ref{lst:cus_do_pages_move}). 
We evaluate the performance of these 3 different migration policies across different workloads and migration loads. 
~\Cref{exp:exp2_intel,exp:exp2_amd} give the performance of a main memory B$^+$-Tree using \texttt{move\_pages2} on a NUMA and a Chiplet architecture, respectively. 
From the experiments on both architectures, it is evident that \texttt{MIGRATE\_SYNC\_LIGHT} is the worst performing migration mode, under-performing by up to 1.3$\times$ compared to the \texttt{MIGRATE\_ASYNC} migration mode, both in terms of B$^+$-Tree query throughput and page migration throughput. Migration modes \texttt{MIGRATE\_ASYNC} and \texttt{MIGRATE\_SYNC} maintain competitive performance for the YCSB-A$^*$ and YCSB-C$^*$ workload in the two architectures. However, in the chiplet architecture, for YCSB-E$^*$ workload, migration mode \texttt{MIGRATE\_ASYNC} outperforms \texttt{MIGRATE\_SYNC} under low migration load by 18\% in terms of B$^+$-Tree query throughput. However, this observation flips when the migration load increases, as \texttt{MIGRATE\_SYNC} achieves 15\% better B$^+$-Tree query throughput over \texttt{MIGRATE\_ASYNC}. 
}


\begin{figure*}[htbp]
    \captionsetup{belowskip=-6pt}
    \begin{subfigure}{0.35\textwidth}
		\centering
	    \includegraphics[width=\columnwidth]{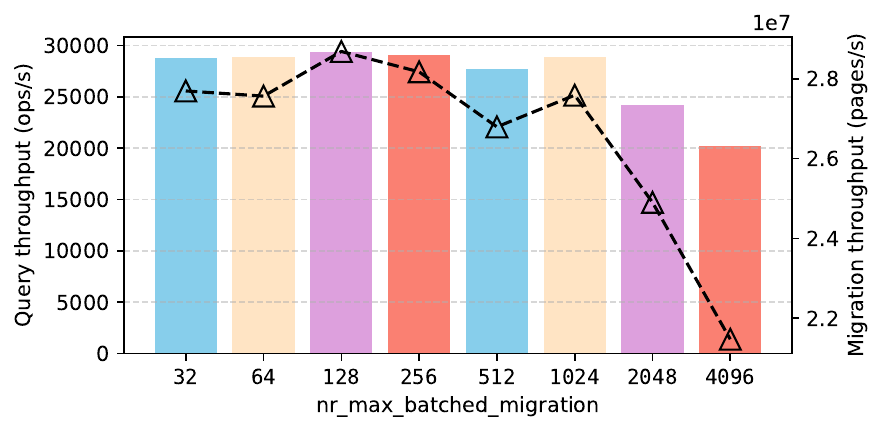}
         \caption{NUMA architecture.}
        \label{exp:exp3_intel}
	\end{subfigure}
	\begin{subfigure}{0.35\textwidth}
		\centering
	    \includegraphics[width=\columnwidth]{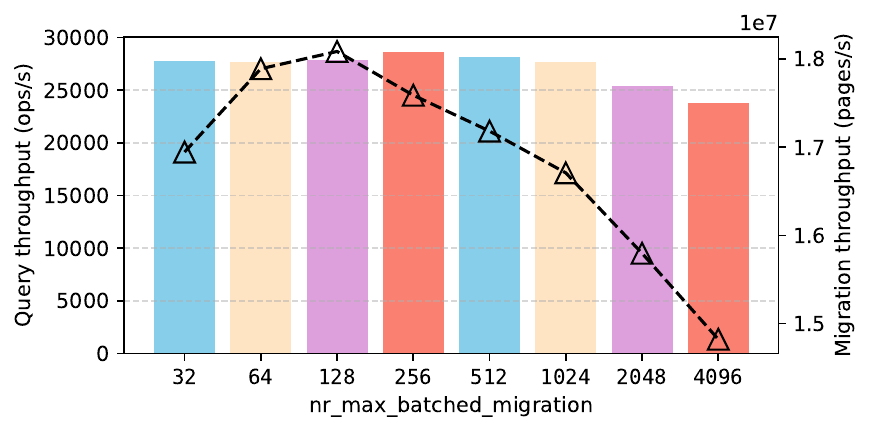}
         \caption{Chiplet architecture.}
        \label{exp:exp3_amd}
	\end{subfigure}
    \caption{Performance of different batch sizes in \texttt{move\_pages2}.}
	\label{exp:exp3}
\end{figure*}


\vspace{7pt}
\noindent\textbf{Maximum Batch Size:} 
{\texttt{move\_pages2} introduces the \vteal{nr\_max\_batched\_migration} knob that lets 
the B$^+$-Tree 
control the number of pages that can be simultaneously migrated during a single invocation of the \texttt{migrate\_pages(2)} kernel function (cf. Line~\ref{mig:cut_1} in Listing~\ref{lst:migrate_pages2}). 
Figure~\ref{exp:exp3} 
gives
the effect of maximum batch size on the performance of a main memory B$^+$-Tree using \texttt{move\_pages2} on both 
the
NUMA and chiplet architectures for the YCSB-C$^*$ workload under low migration load. 
It is evident from Figure~\ref{exp:exp3} that the maximum batch size significantly impacts the page migration performance of the B$^+$-Tree. The default Linux setting, i.e, a maximum batch size of 512 pages, under-performs in both architectures for both B$^+$-Tree query throughput and page migration throughput. As the batch size increases, both the B$^+$-Tree query and page migration throughputs 
improve until they reach a plateau, then the performance starts to degrade. On the NUMA machine, the main memory B$^+$-Tree achieves the best query throughput when the batch size is set to 128, outperforming the default Linux settings by up to 5\%. In the chiplet architecture, the optimal maximum batch size shifts to 256, yielding 1.5\% more query throughput over the default settings. The B$^+$-Tree page migration throughput exhibits even better performance. The optimal batch size provides 7.1\% and 2.4\% better page migration throughput over the default Linux settings on the NUMA and Chiplet architecture, respectively. }

\section{Discussion}
\begin{figure}[htbp]
    \centering
    \includegraphics[width=\columnwidth]{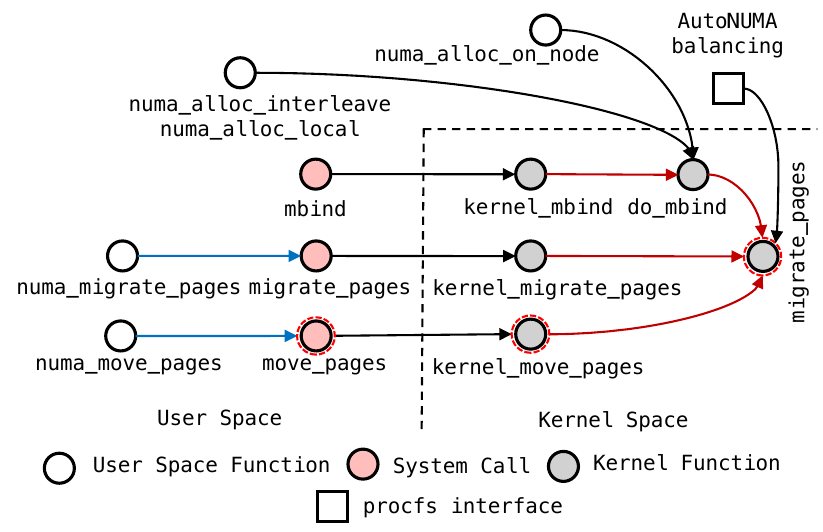}
    \caption{Function call graph of various functions invoking \texttt{move\_pages} and \texttt{migrate\_pages}. The red-bordered circles indicate the functions that have been customized in this work.}
    \label{fig:others}
\end{figure}

\noindent\textbf{System-wide Implications of the Modified System Call:} {Figure~\ref{fig:others} showcases various user functions, kernel functions, and procfs interfaces~\cite{linux_sysfs} that directly or indirectly invoke \texttt{move\_pages} and \texttt{migrate\_pages}, and therefore can benefit from the modifications introduced in this paper.
The widely used Linux package numactl~\cite{linux_numactl} offers an API through the libnuma shared library. This enables applications, e.g., an MMDB to implement NUMA-aware memory management and scheduling. Multiple user-space libnuma functions, e.g., numa\_move\_pages, numa\_migrate\_pages, numa\_alloc\_interleave, numa\_alloc\_local, numa\_alloc\_on\_node call the \texttt{move\_pages} system call and \texttt{migrate\_pages} kernel function. Observe that functions, e.g., numa\_alloc\_interleave, numa\_alloc\_local and numa\_alloc\_on\_node serve as NUMA-aware memory allocators. Aside from the libnuma user-functions, \texttt{mbind}~\cite{mbind} is another widely used system call that indirectly invokes the \texttt{migrate\_pages} kernel function. Linux introduces an automatic 
NUMA
balancing feature, i.e., AutoNUMA~\cite{linux_autonuma} in order to proactively move pages across different sockets to improve data locality. This feature can be enabled by setting the corresponding proc file system entry 
at \texttt{/proc/sys/kernel/numa\_balancing} to 1. Internally, the AutoNUMA feature inherently invokes the \texttt{migrate\_pages} kernel function to complete the page migration procedure. An MMDB leveraging these user functions and system calls can further benefit from the modifications by invoking \texttt{move\_pages2} and \texttt{migrate\_pages2} in place of the respective \texttt{move\_pages} system call and \texttt{migrate\_pages} kernel function. 
}

\vspace{7pt}
\noindent\textbf{Limitations of the Modified System Call:} {The thrust of \texttt{move\_pages} is to alter the physical ownership of a memory page in an application-agnostic manner by keeping the virtual address of the migrated memory page unchanged. As a result, from the MMDB's application point of view, nothing changes, i.e., the MMDB uses the same virtual address to locate the same memory page. It is the cost of accessing the memory page that changes, depending on its physical location in the hardware landscape. The same applies to the modified system call, \texttt{move\_pages2}. The modifications introduced for the custom system call are motivated by the observation that it is possible to mitigate the performance bottlenecks associated with the ``abort-on-failure" strategy of the original \texttt{move\_pages} by allowing partial migration, i.e., migrating as many pages as possible without halting on failure. This approach is effective for MMDBs in scenarios where partial migration is acceptable. Refer to the use cases discussed in Section~\ref{sec:use_case}. These use cases revolve around the notion of moving hot data as close to the compute chip as possible. While failure to migrate a memory page in such cases will impact the MMDB performance, it does not compromise the stability of the database. However, cases where it warrants for a complete migration, i.e., all the memory pages need to be migrated during a migration round, the performance improvement of the proposed system call diminishes. 
}


\vspace{7pt}
\noindent\textbf{Implications of the Modified System Call on the Performance of a Main Memory B$^+$-Tree:} 
Due to the difference in access latencies across different NUMA sockets in a NUMA architecture and chiplets in a Chiplet Architecture, a MMDB may need to migrate B$^+$-Tree index nodes and data across these NUMA sockets and chiplets~\cite{BangOMPB20}. Thus, apart from the B$^+$-Tree query throughput, a high page migration throughput becomes also essential for such hardware architectures. During migration, compared to the Linux baseline, the proposed system call improves the B$^+$-Tree's query throughput by up to 2.2$\times$ and 2.3$\times$ in NUMA and Chiplet architectures, respectively. Likewise, the proposed system call improves the B$^+$-Tree's migration throughput by up to 2.2$\times$ and 2.6$\times$ for the respective architectures. \texttt{move\_pages2} also enables an MMDB to tune the migration behavior of the B$^+$-Tree through knobs \texttt{mode} and \texttt{nr\_max\_batched\_migration}. Fine-tuning these knobs can yield up to 18\% and 7.1\% performance improvement in terms of B$^+$-Tree query throughput.


\section{Concluding Remarks}
{\vpurple{move\_pages} is the workhorse for on demand page migration in modern hardware architectures. In this paper, we propose \vpurple{move\_pages2}, a custom system call for page migration in main-memory databases in the context of modern hardware, upholding the spirit of DB-OS co-design}. This system call mitigates the cost of context switches by migrating as many pages as possible within a single invocation. Additionally, it offers configurable knobs that allow the MMDB to tailor the migration process based on the query workload and migration load. Compared to a main memory B$^+$-Tree implementation using the native page migration system call, the B$^+$-Tree implementation using \vpurple{move\_pages2} can achieve up to 2.3$\times$ higher query throughput, and 2.6$\times$ higher page migration throughput.


\bibliographystyle{ACM-Reference-Format}
\bibliography{sample-base}

\end{document}